\documentclass[pra,showpacs,twocolumn]{revtex4}
\usepackage{graphicx}
\usepackage{amsmath}
\usepackage{amsfonts}
\begin{document}
\title{Threshold for creating excitations in a stirred superfluid ring}
\author{K. C. Wright}
\altaffiliation{Present Address: Department of Physics and Astronomy, Dartmouth College, Hanover, NH, 03755}
\author{R. B. Blakestad}
\altaffiliation{Present Address: Booz Allen Hamilton, Arlington, VA, 22203}
\author{C. J. Lobb}
\altaffiliation{Also at the Center for Nanophysics and Advanced Materials, University of Maryland, College Park, MD, 20742, USA}
\author{W. D. Phillips}
\author{G. K. Campbell}
\affiliation{Joint Quantum Institute, National Institute of Standards and Technology and University of Maryland, Gaithersburg, MD, 20899, USA}


\begin{abstract}
We have measured the threshold for creating long-lived excitations when a toroidal Bose-Einstein condensate is stirred by a rotating (optical) barrier of variable height. When the barrier height is on the order of or greater than half of the chemical potential, the critical barrier velocity at which we observe a change in the circulation state is much less than the speed for sound to propagate around the ring. In this regime we primarily observe discrete jumps (phase slips) from the non-circulating initial state to a simple, well-defined, persistent current state. For lower barrier heights, the critical barrier velocity at which we observe a change in the circulation state is higher, and approaches the effective sound speed for vanishing barrier height. The response of the condensate in this small-barrier regime is more complex, with vortex cores appearing in the bulk of the condensate. We find that the variation of the excitation threshold with barrier height is in qualitative agreement with the predictions of an effective 1D hydrodynamic model.
\end{abstract}

\maketitle

The critical flow velocity of a superfluid is intimately connected with the spectrum of allowed excitations for a quantum fluid, and can provide insight into the mechanisms which produce and sustain superfluidity in that system. This connection between the excitation spectrum and the critical velocity of a superfluid was first identified by Landau~\cite{LandauJPUSSR41}, who showed that there is a minimum velocity above which it becomes energetically possible to create excitations. In an \emph{infinite, homogeneous} dilute Bose-condensed gas, the excitation spectrum is given by the Bogoliubov dispersion relation, the elementary excitations are phonons, and the critical velocity for a pointlike defect is the Bogoliubov speed of sound~\cite{BogoliubovSound, AstrakharchikMotionPRA04}. 

In any \emph{real} superfluid system, the critical velocity can be modified by the finite system size, including reduced dimensionality, and by various inhomogeneities, \emph{e.g.,} surface roughness, inhomogeneous confining potentials, and the size and shape of any moving defects~\cite{FeynmanApplicationPiLTP55, AndersonConsiderationsRMP66, LangerIntrinsicPRL67, SchwarzEffectPRL92, FrischTransitionPRL92, WinieckiPressurePRL99, StiessbergerCritcalPRA00, CrescimannoAnalyticalPRA00, FedichevCriticalPRA01, AstrakharchikMotionPRA04}. Such effects can give rise to dissipation through the creation (and subsequent motion) of elementary excitations such as solitons and vortices~\cite{FeynmanApplicationPiLTP55, AndersonConsiderationsRMP66, LangerIntrinsicPRL67, SchwarzEffectPRL92, FrischTransitionPRL92, WinieckiPressurePRL99, StiessbergerCritcalPRA00, CrescimannoAnalyticalPRA00, FedichevCriticalPRA01}, in addition to phonons. In general, coupling to these other modes of excitation causes the critical velocity for a moving disturbance to be lower than the sound speed. Inhomogeneity and the details of the geometry thus play an important role in the onset of dissipation in a superfluid.

Historically, most experimental studies of the superfluid critical velocity were conducted with liquid helium~\cite{KapitsaViscosityN38, AllenFlowN38, VinenDetectionPotRSoLA61, TrelaSuperfluidPRL67}, including a variety of increasingly sensitive experiments conducted in an annular geometry~\cite{ReppyApplicationPRL65, ClowTemperaturePRL67, BendtSuperfluidPR62, BendtThresholdPRL1967, AvenelDetectionPRL97, SchwabSuperfluidJoLTP98, SimmondsQuantumN01, BrucknerLargeJoAP03, HoskinsonSuperfluidPRB06}. More recently, degenerate quantum gases of neutral atoms have provided new possibilities for studying the superfluid state~\cite{LeggettQuantum06}. The earliest experiments reporting a critical velocity in an ultracold atomic gas were conducted in \emph{simply-connected} Bose-Einstein condensates, where a perturbing potential was moved through the condensate, and the onset of dissipation was detected as heating of the condensate~\cite{RamanEvidencePRL99, OnofrioObservationPRL00, RamanDissipationlessJLTP01}. In related experiments, a threshold for the nucleation of vortices and solitons was observed when a condensate was perturbed by a moving potential defect~\cite{InouyeObservationPRL01,EngelsStationaryPRL07, NeelyObservationPRL10}. Additionally, a critical rotation frequency was observed in experiments where a simply connected condensate was stirred with a rotating potential~\cite{MadisonVortexPRL00, HaljanDrivingPRL01, Abo-ShaeerObservationS01}. Critical velocity measurements have also been undertaken with an ultracold Fermi gas across the BEC-BCS crossover~\cite{MillerCriticalPRL07}, and with a 2D trapped Bose gas~\cite{DesbuquoisSuperfluidNP12}.

\begin{figure}
\includegraphics[width=3.25in]{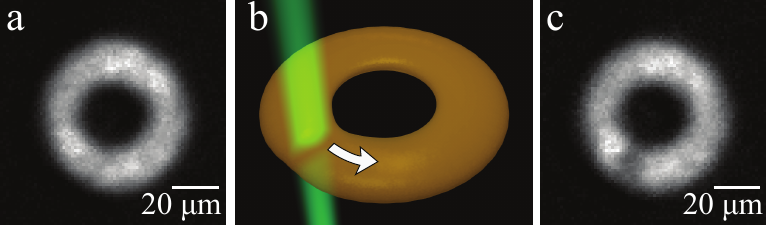}
\caption{(a) \emph{in situ} absorption image of the ring condensate without a barrier, viewed from above. (b) Schematic showing the orientation of the blue-detuned ($\lambda$=532 nm) ``barrier'' beam used to create a barrier in the ring condensate. The arrow indicates the azimuthal movement of the barrier around the ring axis. (c) \emph{in situ} absorption image showing the effect of the barrier beam on the ring condensate. The peak barrier height in (c) is $\approx$ 40\% of the chemical potential. (a) and (c) are the average of 5 partial-transfer absorption images~\cite{RamanathanPartial-TransferRoSI12} of different condensates, with a 96$\times$96 $\mu$m field of view.}
\label{fig:trapgeom}
\end{figure}

Recent experimental successes in creating atomic gases in an \emph{annular} geometry~\cite{RyuObservationPRL07, SherlockTimePRA11, HendersonExperimentalNJP09, RamanathanSuperflowPRL11, MoulderQuantizedPRA12, BeattiePersistentPRL13, WrightDrivingPRL13, MartiCollectiveA13, RyuExperimentalA13} have provided a new opportunity for further studies of the properties of the superfluid state. We previously reported the first measurement of a critical flow velocity in a superfluid ring~\cite{RamanathanSuperflowPRL11} by observing the decay of a persistent current flowing past a \emph{stationary} optical barrier as we varied the barrier height. In~\cite{WrightDrivingPRL13}, we observed discrete phase slips in a ring geometry perturbed by a \emph{moving} barrier. In that work, the barrier was moving at an angular velocity much less than the velocity of sound propagating around the ring. Here, we study the creation of excitations and therefore dissipation over a wider range of conditions using a variable-height barrier with an angular velocity ranging from zero up to the speed of sound. Because our superfluid ring supports long-lived persistent currents, we are able to detect the threshold at which excitations occur with a high degree of sensitivity by measuring changes in the circulation state~\cite{RamanathanSuperflowPRL11, MoulderQuantizedPRA12, BeattiePersistentPRL13, WrightDrivingPRL13}. This experiment was conducted by creating a ring-shaped condensate in a non-circulating state, then stirring it for one second with a small (diameter less than the width of the annulus) repulsive potential (created by a focused blue-detuned laser beam) moving azimuthally at a fixed angular velocity (Fig.~\ref{fig:trapgeom}).  Repeating this procedure many times for various combinations of potential barrier height and angular velocity, we have determined how the threshold for creation of such long-lived excitations depends on these experimental parameters.

In Sec.~\ref{sec:expt} of this paper, we describe the experiment in detail, and report our observations of the threshold for creating excitations in the ring. In Sec.~\ref{sec:theory} we present a 1D hydrodynamic model of our ring condensate, which incorporates elements from the work of Watanabe et al.~\cite{WatanabeCriticalPRA09}, and Fedichev and Shlyapnikov~\cite{FedichevCriticalPRA01}. In Sec.~\ref{sec:discussion} we then compare our data to the model's prediction of the critical barrier height for a given barrier velocity.

\section{Experimental Procedure and Results}\label{sec:expt}

The superfluid ring in our experiments was a Bose-Einstein condensate of $7.6(20) \times 10^5$ $^{23}$Na atoms in the $3^2S_{1/2}\left|F=1,m_F=-1\right>$ state~\cite{endnote:uncertainty}, at a temperature of $< 40$ nK.  The toroidal optical dipole trap for the atoms was created in the same manner as reported by us previously~\cite{RamanathanSuperflowPRL11, WrightDrivingPRL13}, using a horizontally propagating ``sheet'' beam, and a vertically propagating Laguerre-Gauss (LG$_0^1$) ``ring'' beam generated using a phase hologram~\cite{HeckenbergLaserOaQE92}. The wavelength of both beams was $\lambda=1030$ nm, far red-detuned from the $^{23}$Na $D_2$ resonance at $\lambda=589$ nm. Together, these two beams created an attractive dipole trap described (in the harmonic approximation) by
\begin{equation}
U_\mathrm{trap}(\rho,z) = \frac{1}{2}m\left[\omega_z^2 z^2 + \omega_\rho^2(\rho-R)^2 \right],
\label{eq:trap}
\end{equation}
where $m$ is the atomic mass, $\omega_z$ ($\omega_\rho)$ is the trap frequency in the vertical (radial) direction, and $R$ is the radius of the ring. In this experiment the trap parameters were measured to be $\omega_r/2\pi$ = 134(6) Hz, $\omega_z/2\pi$ = 550(20) Hz, and $R$ = 22.6(2.3) $\mu$m. The measured Thomas-Fermi width of the condensate in the $\rho$ (radial) direction was 22(2) $\mu$m. Using the Thomas-Fermi approximation we calculate the chemical potential to be $\mu_0/h$ = 2.1(2) kHz~\cite{endnote:chemicalpotential}. With this chemical potential and the measured vertical trap frequency we calculate that the maximum vertical thickness (Thomas-Fermi) of the ring should be 5.0(1)$\mu$m. This is roughly consistent with what we observe, given the $\approx4$ $\mu$m resolution limit of the horizontal imaging system.

The repulsive barrier potential used to ``stir'' the condensate was created by a blue-detuned ($\lambda=532$ nm) laser beam focused to a circular spot 9(1) $\mu$m in diameter (FWHM), which is smaller than the width of the annulus.  The intensity and position of the beam were controlled by a two-axis acousto-optic deflector (AOD). Stirring beam powers of up to $\approx70$ $\mu$W were used in the experiment, resulting in a peak barrier potential height of $U_b/h=1.30(13)$ kHz. The barrier height was calibrated by measuring the density depletion of the condensate \emph{in situ} as a function of laser power and position. The stated 10\% uncertainty in the barrier height reflects the uncertainty in measuring the density depletion of the condensate at the location of the barrier~\cite{endnote:barriercalib}. The height of the barrier also varied systematically with position around the ring by $\approx10\%$ due to the angle-dependent diffraction efficiency of the AOD and angle-dependent losses in the imaging system.

Because of the azimuthal variation of the barrier height and the variation in condensate density around the ring, the experiment was designed so that the barrier always made at least one full revolution around the ring during the one second stirring procedure. We expect that the creation of excitations primarily occurred at the weakest point, $i.e.$ where the fractional height of the barrier was largest compared to the local maximum value of the interaction energy (the 3 o'clock position in Fig.~\ref{fig:trapgeom}). The comparison of our experimental results to the theory in section \ref{sec:theory} assumes that this is the case.

Prior to stirring the condensate, we ramped on the (stationary) barrier beam in 100 ms to a height sufficient to stop any spuriously formed persistent currents~\cite{endnote:spontaneous}, then ramped the intensity back to zero in another 100 ms. With the condensate in this non-circulating state, we then stirred it with the moving barrier at constant angular velocity $\Omega$ for a total duration of one second. The barrier height was ramped up (while rotating) from zero to a value $U_b$ in 100 ms, held at that height for 800 ms, then ramped off in another 100 ms.

\begin{figure}
\centering
\includegraphics[width=80mm]{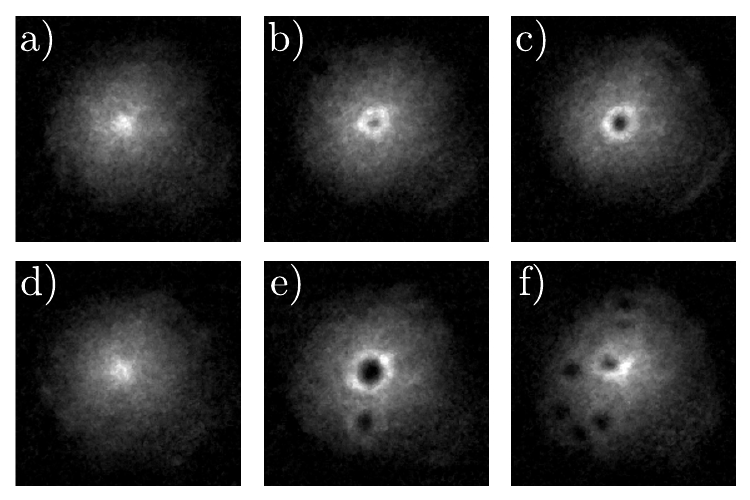}
\caption{Time-of-flight (TOF) absorption images showing vertical column density profiles observed after one second of stirring, adiabatic relaxation of the radial trap confinement (see text), and 10 ms of expansion. The upper row (a-c) shows representative results for stirring at low speeds $\Omega/2\pi<5$ Hz, while the lower row shows results for high speeds $\Omega/2\pi>5$ Hz. For sufficiently small barrier height $U_b$ (a,d), the condensate remains in the non-circulating state, and the density profile is peaked in the center after TOF expansion, with no evidence of vortices. For higher $U_b$ the ring can be excited to a persistent current state (b,c,e,f), causing a central hole to appear in the TOF density profile. The size of the central hole increases with the phase winding number $l$: in (b) $l=1$, in (c) $l=2$, and in (e) $l=3$. At high $\Omega$, the stirring may produce of off-axis vortices, as seen in (e,f). At high $\Omega$ and sufficiently high $U_b$ (f), many vortices appear and the central hole associated with the persistent current may be distorted.  The stirring conditions ($\Omega/2\pi$, $U_b/h$) for each image are: (a) 1 Hz, 1080 Hz  (b) 2 Hz, 930 Hz (c) 2 Hz, 1010 Hz (d) 30 Hz, 40 Hz (e) 8 Hz, 370 Hz (f) 25 Hz, 60 Hz.}
\label{fig:TOFresults}
\end{figure}

After this stirring procedure, we detected the presence of excitations in the condensate using a time-of-flight (TOF) imaging procedure. The evolution of the ring condensate after release is not trivial~\cite{MurrayProbingTBP13}, but can be used to determine the circulation state of the condensate prior to release~\cite{WrightDrivingPRL13, MoulderQuantizedPRA12}. In the absence of any circulation the expansion of the condensate causes the central hole to close, after which the density profile typically exhibits a central peak surrounded by a broad pedestal (Fig.~\ref{fig:TOFresults}[a,d]) when imaged along the (vertical) symmetry axis of the trap. In contrast, when a condensate with some form of circulation is released, the density profile after TOF expansion shows one or more holes due to the presence of phase singularities (vortices) in the condensate wave function (Fig.~\ref{fig:TOFresults}[b,c,e,f]). 

A central hole in the density profile after TOF expansion signifies the presence of a persistent current flowing around the ring. The size of the central hole depends on the phase winding number of the persistent current and the velocity of the mean-field-driven inward expansion. If we release our ring condensate by suddenly and simultaneously turning off both of the trapping beams, the hole is too small to be resolved by our imaging system for experimentally accessible TOFs ($<$ 15 ms). As in our previously reported work~\cite{RamanathanSuperflowPRL11, WrightDrivingPRL13}, we make the signature of circulation visible earlier by first adiabatically reducing the ring beam intensity by 90\% over 100 ms, then releasing the condensate suddenly into ballistic expansion. We used this procedure, followed by 10 ms TOF and partial-transfer~\cite{RamanathanPartial-TransferRoSI12} absorption imaging, to detect excitation of the superfluid ring for all the data presented here. When we follow this procedure we find that the radius of the central hole in TOF increases roughly linearly with the winding number of the persistent current~\cite{MurrayProbingTBP13}.

In addition to the central hole, which signifies a persistent current, in some cases we also observed off-axis holes in the density profile after TOF (Fig.~\ref{fig:TOFresults}[e,f]), indicating the presence of vortex excitations in the annulus. While we note that there was a higher probability of observing such off-axis excitations with a barrier moving at higher angular velocities, we did not separately analyze and quantify the probability of observing them. For this experiment, the appearance of one or more ``holes'' in the density profile of the condensate after TOF was construed as evidence that the threshold for creating excitations had been exceeded.

\begin{figure}
\centering
\includegraphics[width=85mm]{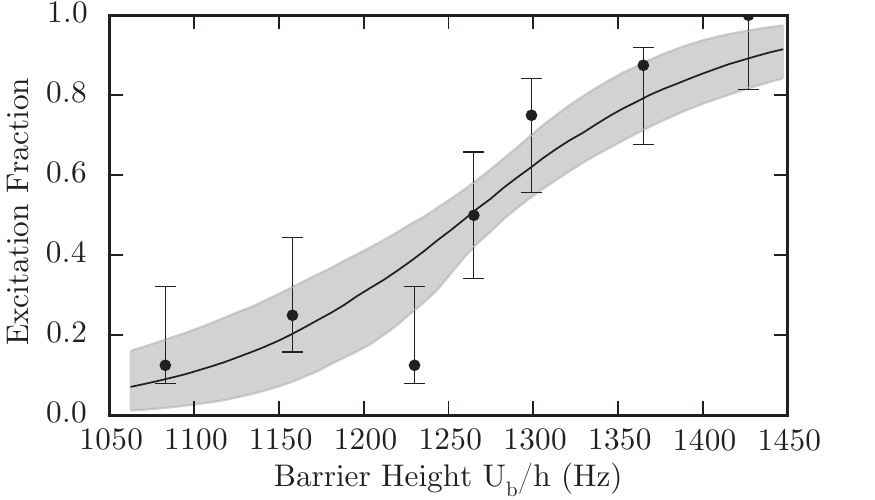}
\caption{Fraction of experimental runs where an excitation was observed, as a function of the peak height of the potential barrier $U_b$, with the barrier moving at an angular velocity $\Omega/2\pi$ = 1 Hz. The experiment was repeated 6-8 times for each value of $U_b$ (black dots). The vertical error bars are the statistical uncertainty in the measured excitation fraction for each barrier height $U_b$.  The black curve is a least-squares fit of a sigmoidal function (Eq.~\eqref{eq:sigmoid}) to the data points. The gray region is a 68\% confidence band for the sigmoidal fit (see text).}
\label{fig:sigfits}
\end{figure}

Using this criterion, we determined the probability of excitation for 10 values of $\Omega$, with $\Omega/2\pi$ ranging from 1 to 30 Hz. For each $\Omega$, the probability of excitation was found for a wide range of $U_b$ by conducting the experiment repeatedly at each specific value of $\Omega$ and $U_b$, then varying $U_b$ until we had mapped out a range over which the probability of excitation changed from nearly zero to nearly unity, as shown in Fig.~\ref{fig:sigfits}. The highest value of $\Omega/2\pi$ (30 Hz) is close to the angular velocity at which sound is expected to propagate around the ring (see section~\ref{sec:theory}).

\begin{figure}
\centering
\includegraphics[width=85mm]{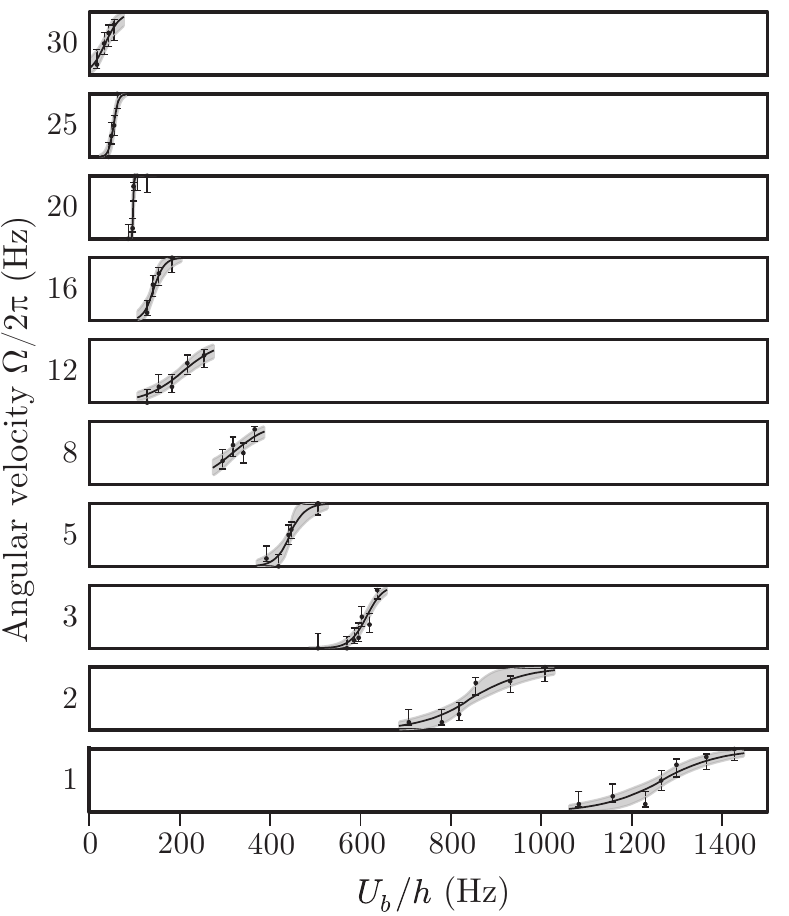}
\caption{Comparison of data for each of the selected values of $\Omega$ used in the experiment, as plotted in Fig~\ref{fig:sigfits}, with barrier height $U_b$ on the horizontal axis. The value of $\Omega/2\pi$ in Hertz for each plot is shown on the left. The black dots are the observed excitation fraction (ranging from zero to one for each sub-plot) for a given value of $U_b$ and $\Omega$, each dot representing 6-8 repetitions.  The vertical error bars are the statistical uncertainty in the measured excitation fraction for each barrier height $U_b$. The black curves are least-squares fits of a sigmoidal function (Eq.~\eqref{eq:sigmoid}) to the black dots. The gray regions are 68\% confidence bands for the sigmoidal fits (see text).}
\label{fig:allplots}
\end{figure}

In analyzing the data, the fraction of excitations observed in repeated experiments at each value of $U_b$ and $\Omega$ was used as an estimate of the true probability of excitation. In order to estimate uncertainties, we assume that the probability distribution for excitation is binomial, and use the beta distribution as an approximation to the discrete binomial distribution, following the approach of Ref.~\cite{CameronEstimationPASA11}. The error bars shown in Figs.~\ref{fig:sigfits} and \ref{fig:allplots} are the 68\% confidence interval, as estimated from the beta distribution. We took the critical barrier height $U_c$ for a given angular velocity $\Omega$ to be the value of $U_b$ at which there is a 50\% probability of observing an excitation in the ring condensate after the experimental stirring procedure. These values were determined from the data for each $\Omega$ by a least-squares fit of the sigmoidal function 
\begin{equation}
P(U_b) = 1/(1+e^{(U_c-U_b)/\delta U})\label{eq:sigmoid}
\end{equation}
to the measured probability of excitation at each $U_b$ (Fig.~\ref{fig:sigfits}), where each point was weighted by the number of samples ($\delta U$ is the width of the sigmoidal fit). To estimate the statistical uncertainty in this fit, we employed a parametric bootstrapping~\cite{EfronIntroduction93} method. The observed probability of excitation and number of samples at each point were used to specify beta-distributed random variables associated with those points. Samples were then drawn from these distributions, using the same sample sizes as in the original data set. This simulated data was then fit using Eq.~\eqref{eq:sigmoid}. This procedure was repeated 1000 times for each value of $\Omega$, and the set of all simulated fits was used to estimate the $1\sigma$ uncertainty in the measurement of the critical barrier height for each $\Omega$. The confidence bands displayed in Figs.~\ref{fig:sigfits} and~\ref{fig:allplots} are the 15.9 (lower) and 84.1 (upper) percentiles of the excitation probabilities calculated from the set of all fits to simulated data for a given $\Omega$. We note that the parameters $U_c$ and $\delta U$ for each set of simulated fits were not always normally distributed.

Figure~\ref{fig:allplots} displays the data for each of the selected values of $\Omega$ used in the experiment, showing the character of the data and the sigmoidal fit for each value of $\Omega$. As expected, for low angular velocities, the barrier height must be a large fraction of the chemical potential before excitations occur, and this critical barrier height decreases as the angular velocity increases. For low $\Omega$, the excitations generally occur as simple phase slips from the non-circulating state to a persistent current state with a phase winding number $l=1$, with no indication of vortices within the annulus [Fig~\ref{fig:TOFresults}(b)]. As we showed previously~\cite{WrightDrivingPRL13}, in this regime the weak link created by the rotating barrier can act like a Josephson junction, in that it allows quantized jumps in the persistent current state of the superfluid in the ring. While we note that the widths of our fits to the data vary as a function of $\Omega$, we believe that more data sampling and better control of experimental conditions would be required to draw any detailed conclusions about these widths.

At higher $\Omega$ the response of the condensate can become more complex. Excitations appear for much smaller barrier heights, but in this case we more frequently observe vortices within the annulus [Fig.~\ref{fig:TOFresults}(e,f)] that have survived to be observed in TOF~\cite{endnote:vortexlifetime}. When phase slips to different persistent current states did occur, it was almost always to states with winding number $l$=1 or 2, even when the angular velocity was much higher than the rotation rate associated with one quantum of circulation in the ring $\Omega_0/2\pi\approx \hbar/mR^2$ = 0.86 Hz. Transitions to higher circulation states ($l>2$) [Fig.~\ref{fig:TOFresults}(e)] typically only occurred when the barrier height was well above the critical barrier height. When the first excitation occurs, the fact that the system does not necessarily relax to the global state of lowest energy (in the rotating frame), but instead typically settles into the first energy minimum where the flow velocity through the barrier is less than the critical velocity may indicate that the phase slip dynamics in this system are strongly damped~\cite{DuzerPrinciples98}.

Extracting the value of $U_c$ from each of these fits gives us information about the critical behavior of the system. In Fig.~\ref{fig:criticalvelocity} we plot $U_c$ as a function of $\Omega$, and compare it to the predictions of an effective 1D theoretical model presented in the next section.

\section{Theoretical Model}\label{sec:theory}

While a detailed understanding of the dynamics in our experimental system may require a full 3D model~\cite{PiazzaVortex-inducedPRA09, PiazzaInstabilityNJoP11}, the basic features of our data can be described by an effective 1D model of the system. To create this model, we treat flow around the ring as if in a (locally) straight channel with a single potential energy barrier of height $U_b$, and neglect the periodic boundary conditions. We assume that in our experiment the temperature ($T<40$ nK) is close enough to zero that the condensate can be described accurately by the Gross-Pitaevskii equation~\cite{PitaevskiiBose-Einstein03}. Furthermore, the smallest features of the trapping potential in our experiment are large compared to the condensate healing length ($\xi\approx$ 0.3 $\mu$m), allowing us to make
the local density approximation, and treat the condensate as locally homogeneous. 

The condensate in our experiments interacts with a smoothly varying (not hard-walled) optical dipole potential that we approximate as
\begin{equation}
U(\vec{r},t)=U_\mathrm{trap}(\rho,z) + U_b(\theta,t),
\end{equation}
where $U_\mathrm{trap}(\rho,z)$ is given by Eq.~\eqref{eq:trap}, and $U_b(\theta,t)$ is a barrier potential with a maximum $U_b$ at an some angle $\theta_b$, and which moves at a constant angular velocity $\Omega$.  We assume for simplicity in the model that the barrier height is independent of the transverse coordinates $\rho$ and $z$. The real barrier potential is $\rho$ dependent, and we note that averaging over the transverse degrees of freedom in this way is less accurate for large barrier heights, at which the radial profile is significantly modified. In the model the trap potential has no dependence on $\theta$, and we can remove the time dependence in the problem by switching to a reference frame co-rotating with the barrier at an angular velocity $\Omega$. From here forward all expressions are time-independent and the rotating frame is implicit.

To formulate a description of the steady-state behavior of the condensate, we first note that in the local density approximation, the interaction energy $\mu(\vec{r})$ in a 3D Bose gas is $\mu(\vec{r}) = g n(\vec{r})$~\cite{PitaevskiiBose-Einstein03}, where $n(\vec{r})$ is the density, and $g = 4\pi a_s \hbar^2 /m$ is the strength of the contact interactions between the atoms ($a_s$ is the atomic s-wave scattering length). In the Thomas-Fermi limit, with no flow, the density profile $n(\vec{r})$ of the condensate is
\begin{equation}
n(\vec{r}) = \left(\mu_0-\frac{m}{2}\left[\omega_z^2 z^2 + \omega_\rho^2(\rho-R)^2 \right]-U_b(\theta)\right)/g.
\end{equation}

If there is nonzero flow in the condensate, conservation of energy requires that the velocity field and density profile satisfy a Bernoulli equation at each point $\vec{r}$
\begin{equation}
\mu_I=\frac{1}{2}mv^2(\vec{r})+g n(\vec{r})+U(\vec{r}),
\label{eq:bernoulli}
\end{equation} 
where $\mu_I$ is the chemical potential for steady-state current flow $I$ through the channel. The current is related to the velocity field and density profile by the continuity condition
\begin{equation}
I=\int v(\vec{r})\,n(\vec{r})\label{eq:continuity}\; dA_\perp,
\end{equation}
where the integral is over the channel cross section $A_{\perp}$. 

To make it possible to analytically determine the critical barrier height for a given barrier velocity, we reduce Eqns. \eqref{eq:bernoulli} and \eqref{eq:continuity} to an effective 1D form. To simplify Eq.  \eqref{eq:bernoulli}, consider the form of the equation when we set $\rho=R$ and $z=0$: 
\begin{equation}
\mu_I=\frac{1}{2}mv^2(\theta)+g n(\theta)+U_b(\theta),\label{eq:1Dbernoulli}
\end{equation}
where $n(\theta)$, $v(\theta)$ and $U_b(\theta)$ are the values of these quantities along the center of the channel. If the azimuthal size of the barrier potential is small compared to the ring circumference, $\mu_I$ is independent of $U_b(\theta)$, and the flow velocity and density far from the barrier ($v_\infty, n_\infty$) are related to the flow velocity and density at the barrier peak ($v_b$, $n_b$) by: 
\begin{equation}
\frac{m}{2}v_\infty^2+g n_\infty = \frac{m}{2}v_b^2+g n_b+U_b,\label{eq:atbarrier}
\end{equation}
where $U_b$ is the peak height of the barrier potential, and all quantities are taken to be at the center of the channel. In the limit that we can neglect the effect of periodic boundary conditions on flow around the ring, $v_{\infty}$ is simply identified as the velocity of the barrier along the channel.

\begin{figure}
\centering
\includegraphics[width=85mm]{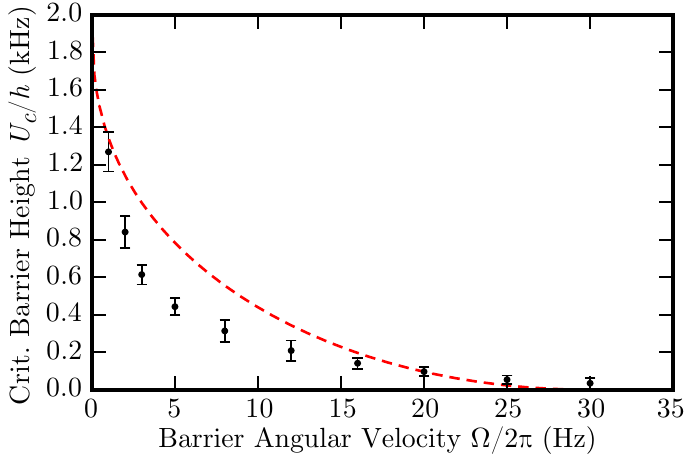}
\caption{Critical barrier height $U_c$ as a function of angular velocity of the rotating potential barrier. The black circles are the values of $U_c$ extracted from fits to the data sets of Fig.~\ref{fig:allplots} at each angular speed $\Omega$. The error bars are the combined statistical uncertainty of the fits and the calibration of the barrier height. The dashed (red) line is the prediction for harmonic confinement in the transverse direction, using Eq.~\eqref{eq:halfpoly}. The normalized theory curve was calculated using $\mu_\infty$ = $g n_\infty$ = 1.98(25) kHz, which assumes excitations were created at the least dense azimuth around the ring. This value of $\mu_\infty$ gives $c^*_\infty = \sqrt{gn_\infty/2 m}$ = 4.24(24) mm/s, which corresponds to an angular velocity $c^*_\infty/2\pi R$ = 30.0(1.7) Hz.}
\label{fig:criticalvelocity}
\end{figure}

Applying the Landau criterion, we may expect that the flow will become dissipative if the superfluid velocity exceeds the local sound speed at the barrier~\cite{LandauFluid87}, since the density and sound speed are at their lowest there, and the flow velocity is highest. Therefore we wish to determine from Eq.~\eqref{eq:atbarrier} the value of $v_{\infty}=\Omega R$~\cite{endnote:periodicBC} for which $v_b$  equals the speed of sound. The speed of sound in a uniform superfluid fluid at $T \approx 0$ is c =$\sqrt{g n/m}$, however, the superfluid flow in our channel has an inhomogenous (2D parabolic) density profile. Hydrodynamic calculations \cite{KavoulakisQuasiPRA98, StringariDynamicsPRA98, ZarembaSoundPRA98} predict that the effective speed of sound in a 2D harmonic channel is reduced by a factor of $\sqrt{2}$ compared to the sound speed calculated using the density at the center of the channel. This can be understood quite simply as the result of the average density in the channel being 1/2 the value at the center of the channel. For our condensate the sound speed at a position $\theta$ is
\begin{equation}
c^*(\theta) = \frac{c(\theta)}{\sqrt{2}}= \sqrt{\frac{g n(\theta)}{2m}}.\label{eq:soundspeed}
\end{equation}
We can thus define the critical barrier velocity $v_c$ and barrier height $U_c$ to be the values of $v_\infty$ and $U_b$ at which
\begin{equation}
v_b = c_b^* = \sqrt{\frac{g n_b}{2 m}}.\label{eq:landauatbarrier}
\end{equation}
This criterion can be used to eliminate $v_b$ from~\eqref{eq:atbarrier}, yielding a relation between $v_c$ and $U_c$:
\begin{equation}
\frac{m}{2}v_c^2+g n_\infty = \frac{5}{4} g n_b + U_c. \label{eq:atbarrier2}
\end{equation}
To eliminate $n_b$ from Eq \eqref{eq:atbarrier2}, we use Eq.  \eqref{eq:continuity} to relate $n$ and $v$ at the barrier and far from it.  To simplify Eq.~\eqref{eq:continuity} we can approximate $v(\vec{r})$ with its value along the center of the channel, $v(\theta)$, and evaluate the integral over the density profile to give
\begin{equation}
I=v(\theta)\int n(\vec{r})\; dA_\perp = v(\theta)\eta(\theta)
\end{equation}
Where the 1D density $\eta(\theta)$ is related to the density along the center of the channel $n(\theta)$ by
\begin{equation}
\eta(\theta)=\frac{\pi g}{m \omega_\rho \omega_z}n^2(\theta).
\end{equation}
In the steady state, $I$ is constant everywhere, and we therefore have the relation
\begin{equation}
n_b^2 v_b=n_\infty^2 v_c \label{eq:continuity2}
\end{equation}
Combining Eq.  \eqref{eq:continuity2} with Eq.  \eqref{eq:landauatbarrier} allows us to derive an expression for $n_b$ in terms of $n_\infty$ and $v_c$:
\begin{equation}
n_b=n_\infty\left(\frac{2 m v_c^2}{g n_\infty}\right)^{1/5} = \\
n_\infty\left(\frac{v_c}{c^*_\infty}\right)^{1/5}\label{eq:nbreplace},
\end{equation}
where $c^*_\infty = \sqrt{g n_\infty/2 m}$ is the effective sound speed far from the barrier. Substituting \eqref{eq:nbreplace} into \eqref{eq:atbarrier2} gives a relation between the critical barrier height $U_c$ and the critical barrier velocity $v_c$ for the condensate under conditions of 2D harmonic confinement,
\begin{equation}
\frac{U_c}{\mu_\infty} = 1 + \frac{1}{4}\left(\frac{v_c}{c^*_\infty}\right)^2 - \frac{5}{4} \left(\frac{v_c}{c^*_\infty}\right)^{2/5} \label{eq:halfpoly},
\end{equation}
where $\mu_\infty = g n_\infty$. The dashed (red) line in Fig.~\ref{fig:criticalvelocity} shows this result plotted against the experimental data. In this plot we have converted barrier velocity along the channel to angular velocity around the ring using $\Omega = v/R$, and used $\mu_\infty$ = 1.98(25) kHz and $c^*_\infty/2\pi R$ = 30.0(1.7) Hz, which are the interaction energy and effective sound speed at the least dense azimuth of the superfluid ring.

\section{Discussion}\label{sec:discussion}

It is clear from Fig.~\ref{fig:criticalvelocity} that the data agree qualitatively with the model (there are no adjustable parameters), but the agreement is not perfect. The data generally falls below the theory for low values of $\Omega$, and is slightly above the theory for sufficiently high values of $\Omega$. Eq.~\eqref{eq:halfpoly} predicts that for vanishing barrier height the critical (angular) velocity of the barrier should asymptotically approach that of the effective sound speed around the ring, $c^*_\infty/2\pi R = 30.0(1.7) $ Hz. Our data appears to approach a value slightly larger than this, but lower than $c_\infty/2\pi R$=42.5 Hz. In this regard the data do roughly support the prediction in Refs.~\cite{KavoulakisQuasiPRA98, StringariDynamicsPRA98, ZarembaSoundPRA98} that the propagation speed of long-wavelength sound in a superfluid channel is reduced in proportion to the average density of the condensate over its cross section.

There are a variety of possible explanations for remaining discrepancies between the model and our data.  We expect some inaccuracy in the model due to our neglect of the transverse variation of the barrier potential, especially at large $U_b$ (small $\Omega$). The fact that the data in general lie below the theory curve may also indicate a critical flow velocity that is smaller than the effective sound speed we assumed in Eq.~\eqref{eq:soundspeed}. It is possible that the actual critical flow velocity for certain excitations, such as vortices, may be less than the sound speed~\cite{FeynmanApplicationPiLTP55, DonnellyStabilityPRL66}. A numerical analysis of the normal modes of a condensate in a 2D harmonic channel by Fedichev and Shlyapnikov~\cite{FedichevCriticalPRA01} has predicted that for clouds with large Thomas-Fermi parameter ($\mu_0/\omega_{\mathrm{trap}}$), the critical flow velocity should be even lower than the effective sound speed used in our derivation of Eq.~\eqref{eq:halfpoly}. Factors such as these may account for part of the discrepancy between the data and our model.

In the high-velocity (low-barrier) regime, the data appear to be somewhat above the value predicted by the model, with a slightly flatter slope. In this regime, the rotating barrier may couple most effectively to radial excitations such as surface waves localized to the inner and outer edges of the ring. Excitation of these modes is not accounted for in our model. The existence of a distinct ``surface'' critical velocity is well established by experiments with condensates in simply-connected geometries~\cite{MadisonVortexPRL00, Abo-ShaeerObservationS01, MadisonStationaryPRL01, RamanVortexPRL01, ZwierleinVorticesN05}. Recent theoretical work has begun to explore the role of surface wave excitations in a toroidal geometry, and indicates that surface waves do play an important role in determining the stability of flow in a superfluid ring~\cite{DubessyCriticalPRA12, WooVortexPRA12}. Finally, the difference between the experiment and theory in this regime could also be due to the fact that we only detect long-lived topological excitations such as vortices and persistent currents, and are not sensitive to other dissipative excitations such as surface waves. Complementary detection techniques such as temperature measurements after the stirring process~\cite{DesbuquoisSuperfluidNP12}, or in-situ detection of surface waves and phononic excitations~\cite{MartiCollectiveA13} may help to distinguish between the role of these different dynamical mechanisms.

\section*{SUMMARY}

We have measured the critical barrier velocity and barrier height for creating excitations when a potential barrier is rotated around a superfluid ring. The experimental data is in qualitative agreement with a 1D hydrodynamic model of flow in the ring. The discrepancies may be due to several causes as discussed above. Thermal fluctuations may also play an important role in lowering the critical barrier velocity below the value predicted by zero-temperature theory~\cite{MatheyDecaya12}. Finally, we have treated our ring geometry as if it were a straight channel, as was done in Ref.~\cite{WatanabeCriticalPRA09}. This omits features such as periodic boundary conditions, quantization of circulation, and curvature of the channel. These present interesting opportunities for further investigation, and may shed further light on the mechanisms that produce and sustain the superfluid state.

\section*{ACKNOWLEDGMENTS}

The authors thank L. Mathey, A. Mathey, C. Clark, M. Edwards, and S. Eckel for insightful discussions, as well as J. G. Lee and Y. Shyur for technical assistance. This work was partially supported by ONR, the ARO atomtronics MURI, and the NSF PFC at JQI. CJL acknowledges support from the NIST-ARRA program.


\begin{thebibliography}{73}
\bibitem[{end({\natexlab{a}})}]{endnote:uncertainty}
\bibinfo{note}{Uncertainties herein are the uncorrelated combination of
  1$\sigma$ statistical and systematic uncertainties unless stated otherwise.}
\bibitem[{end({\natexlab{b}})}]{endnote:chemicalpotential}
\bibinfo{note}{The chemical potential determined by 1) observing the vertical
  expansion of the condensate when released from the trap and 2) calculation
  from the measured number of atoms in the cloud and the trap frequencies were
  different from each other by less than the combined uncertainty of the
  measurements (12\%).}
\bibitem[{end({\natexlab{c}})}]{endnote:barriercalib}
\bibinfo{note}{The maximum barrier height used for stirring calculated using
  the spot size and maximum beam power is 1.2(2) kHz, which is in good
  agreement with the calibration based on the density depletion.}
\bibitem[{end({\natexlab{d}})}]{endnote:spontaneous}
\bibinfo{note}{During evaporative cooling, fluctuations can give rise to
  circulation in the ring. To avoid this, we condense the atoms in the
  (simply-connected) sheet trap, then adiabatically deform the condensate into
  a ring shape. The barrier beam power used to stop any remaining spontaneous
  currents was 100 $\mu$W ($U_b/2\pi$ = 1.9(1) kHz). The measured probability
  for creating an initial state of non-zero circulation with this procedure was
  $<10\%$.}
\bibitem[{end({\natexlab{e}})}]{endnote:vortexlifetime}
\bibinfo{note}{We have separately measured the lifetime of annular vortices
  under similar conditions, and found it to be $\approx$3 seconds.}
\bibitem[{end({\natexlab{f}})}]{endnote:periodicBC}
\bibinfo{note}{If we account for periodic boundary conditions, this relation is
  not strictly true, but the correction is small.}
\end{thebibliography}
\end{document}